\documentclass[a4]{paper}
\usepackage[utf8]{inputenc}
\usepackage[T2A]{fontenc}
\usepackage[english]{babel}
\usepackage{amssymb}
\usepackage{amsmath}
\usepackage{textcomp}
\usepackage{indentfirst}
\usepackage{float}
\usepackage{caption}
\usepackage{textcomp}
\DeclareCaptionLabelSeparator{pipe}{ $: $ }
\usepackage{authblk}

\usepackage{units}
\usepackage{amsfonts}
\usepackage{graphicx} 
\usepackage{amstext} 
\usepackage{wasysym} 

\usepackage{comment}
\usepackage{color}

\usepackage{epstopdf}
\usepackage{xcolor}
\usepackage{hyperref}
 
\definecolor{linkcolor}{HTML}{000000} 
\definecolor{urlcolor}{HTML}{000000} 

\numberwithin{equation}{section}

\hypersetup{pdfstartview=FitH,  citecolor=black, linkcolor=linkcolor,urlcolor=urlcolor, colorlinks=true}
\usepackage[left=2.5cm,right=2cm, top=2cm,bottom=2cm,bindingoffset=0cm]{geometry} 

\usepackage[sorting=none,backend=bibtex,citestyle=numeric]{biblatex}
\bibliography{literature.bib}
 
\begin{document}

\title{Critical Point from Shock Waves Solution in Relativistic Anisotropic Hydrodynamics}

\author{Aleksandr Kovalenko}
\affil{P.N. Lebedev Physical Institute, Moscow, Russia}

\maketitle

\begin{abstract}
Solutions of shock waves in anisotropic relativistic hydrodynamics in the absence of refraction of the flow passing through the shock wave are considered. The existence of a critical value of the anisotropy parameter is shown. This value is the upper limit below which an adequate description of shock waves is possible. Shock wave solutions also provide a mechanism for system isotropization.
\end{abstract}

\section{Introduction}

Description of the evolution of expanding dense and hot hadronic matter using relativistic anisotropic hydrodynamics approach is promising in terms of modeling experimental data \cite{Mubarak2017, Alqahtani2018} and large pressure anisotropies \cite{Strickland, Florkowski2013}, that appears at the early stages of heavy-ion collisions due to the rapid longitudinal expansion. The large difference between the longitudinal and transverse pressures leads to the necessary consideration of high-order gradients in dissipative hydrodynamic theories \cite{Martinez2009}. In relativistic anisotropic hydrodynamics anisotropy is introduced explicitly as an appropriate parameter in one-particle distribution function. The introduction of this parameter is a kind of resummation of the expansion gradients of the theory in a special way, which can give interesting results.

One of the main applications of fluid dynamics is the description of sound propagation and related effects. Sound phenomena in quark-gluon and nuclear matter have been studied mainly in the context of the formation of shock waves \autocite{Rischke1995, Dumitru1999}. The jet-quenching phenomena induce interest in considering the generation of the Mach cone \cite{Satarov2005, CasalderreySolana2007}. Early work has shown that transverse shock waves in hot quark-gluon matter can be generated by initial fluctuations in local energy density (hot spots), which are the result of a large number of QCD interactions \autocite{Gyulassy1996}.

The problem in dissipative hydrodynamic theories is the inability to adequately describe the shock waves phenomenon. For the Israel-Stewart theory, shock waves can be generated only for small Mach numbers \cite{Olson1990, MajoranaMotta}. However, it is possible to obtain discontinuous solutions of shock waves in relativistic anisotropic hydrodynamics by analogy with the isotropic case of an ideal fluid \cite{Landau}. Previously, analytical solutions in the ultrarelativistic case for the longitudinal and transverse shock waves were obtained, and numerical calculations for an arbitrary polar angle were presented \cite{Kovalenko2022}. The main assumption was the introduction of a constant anisotropy. In this regime the anisotropy does not change for the flow moving through the shock wave. For the polar angle $0 < \alpha < \pi/2$ this behavior entails the effects associated with the defection of the flow ($\alpha' \neq \alpha$), as well as the acceleration of the flow for certain values of $\sigma = P^{'}/P$ and $\xi$. Such effects may raise certain questions in practice. Since it is natural to expect the isotropization process for the evolution of hot hadronic matter, the assumption $\xi' = \xi$  may cause the loss of information about the evolution of matter.

Therefore, instead of fixing the anisotropy parameter, one can assume $\alpha' = \alpha$ for the shock waves solutions, which brings us back to the more familiar behavior of the downstream and upstream flows in isotropic case. The present paper describes the derivation of shock wave solutions for such a case. The restrictions on the anisotropy parameter for the existence of shock waves and the system isotropization mechanism are discussed.

\section{Basic equations}

The framework of anisotropic hydrodynamics based on the kinetic theory approach  \cite{MartStr, StrRom1, StrRom2}, where a ansatz for the distribution function $f$ has Romatschke-Strickland form
\begin{equation}
f(x,p) = f_{\textrm{iso}}\Bigg( \frac{\sqrt{p^\mu \Xi_{\mu\nu}(x) p^\nu}}{\Lambda(x)}\Bigg),
\label{Kin1}
\end{equation} 
where $\Lambda(x)$ is a temperature-like momentum scale and $\Xi_{\mu\nu}(x)$ - momentum anisotropy tensor. We consider one-dimensional anisotropy such that $ p^\mu \Xi_{\mu\nu}p^\nu = \mathbf{p}^2 +\xi(x) p_\parallel^2$  in the local rest frame (LRF), where $\xi(x)$ - anisotropy parameter. 

The energy-momentum tensor $T^{\mu\nu}$ can be described in terms of four-velocity vector $U^\mu$ and space-like longitudinal vector $Z^\mu$ as follows
\begin{equation}
T^{\mu\nu} = (\varepsilon + P_\perp)U^\mu U^\nu - P_\perp g^{\mu\nu} + (P_\parallel - P_\perp)Z^\mu Z^\nu,
\label{T_true}
\end{equation}
where $P_\parallel$ and  $P_\perp$ is longitudinal (towards anisotropy direction) and transverse pressure respectively and
\begin{align}
U^\mu &= (u_0 \cosh \vartheta, u_x, u_y, u_0 \sinh \vartheta),
\label{u_aniso}
\\
Z^\mu &= (\sinh \vartheta, 0,0, \cosh \vartheta).
\label{z_aniso}
\end{align}
Here $\vartheta$ is longitudinal rapidity, $u_x, u_y$ -  transverse velocities and $u_0 = \sqrt{1+u_x^2+u_y^2}$.

It is important to note that the dependence on the anisotropy parameter $\xi$ can be factorized:
\begin{align}
\varepsilon & = R(\xi) \varepsilon_{\textrm{iso}} (\Lambda),
\label{e} \\
P_{\perp, \parallel} &= R_{\perp, \parallel} (\xi) P_{\textrm{iso}} (\Lambda),
\label{pTpL}
\end{align}
where the anisotropy-dependent factors $R_\perp(\xi)$ and $R_\parallel(\xi)$ are \cite{MartStr}
\begin{align}
\label{RTL}   
R_\perp(\xi) = \frac{3}{2\xi} \Bigg( \frac{1 + (\xi^2-1)R(\xi)}{1+\xi}\Bigg), \ \ \ \ R_\parallel(\xi) = \frac{3}{\xi} \Bigg( \frac{(\xi+1)R(\xi) -1}{1+\xi}\Bigg),
\end{align}
\begin{align}
R(\xi) = \frac{1}{2} \Bigg( \frac{1}{1+\xi} + \frac{\arctan \sqrt{\xi}}{\sqrt{\xi}}\Bigg). 
\label{R}
\end{align}
Equation of state for the massless gas $\varepsilon_{\textrm{iso}} = 3  P_{\textrm{iso}}$ leads to the following relation between the anisotropic functions: $2 R_\perp(\xi) + R_\parallel(\xi) = 3 R(\xi)$.

We focus on the shock wave solution in the ideal fluid characterized by the energy-momentum tensor (\ref{T_true}). The shock wave in zero-order hydrodynamics is described by a discontinuous solution of the equations of motion \cite{Landau, Israel, Mitchell}.

The energy-momentum conservation leads to the following matching condition linking downstream and upstream projections of  energy-momentum tensor on the direction perpendicular to the discontinuity surface:
\begin{equation}
T_{\mu\nu} N^\mu = T^{'}_{\mu\nu} N^\mu,
\label{gap}
\end{equation}
where \(N^\mu\) - unit  vector normal to the discontinuity surface and  \(T_{\mu\nu}, \ T^{'}_{\mu\nu}\) correspond to upstream and downstream energy-momentum tensors correspondingly.  

Consider a flow moving at an angle \(\alpha\) to the direction of the beam axis. In the case of a normal shock waves for the components of the vector normal to the discontinuity surface we have $N_\mu = (0, \sin \alpha,  0, \cos \alpha)$, where $\alpha$ is polar angle. As discussed earlier, it is assumed that $\alpha' = \alpha$  for the shock waves solutions. Then we have three equations for $v, v', \xi'$ with input parameters $\sigma, \alpha, \xi$.

The equations (\ref{gap}) gives the following system
\begin{align}
\Bigg[ \frac{R_1(\xi)}{1 - v^2} +  \frac{R_2(\xi)}{1 - v^2\cos^2 \alpha} \cos^2 \alpha \Bigg] v - \Bigg[ \frac{R_1(\xi')}{1 - v'^2} +  \frac{R_2(\xi')}{1 - v'^2\cos^2 \alpha} \cos^2 \alpha \Bigg] \sigma v' & = 0,
\label{xi_eq_1} 
\\
\Bigg[ R_\perp(\xi) - \sigma R_\perp(\xi') + \frac{R_1(\xi) v^2}{1 - v^2} -  \sigma \frac{R_1(\xi') v'^2}{1 - v'^2}  \Bigg] \sin \alpha & = 0,
\label{xi_eq_2} 
\\
\Bigg[ R_\perp(\xi) - \sigma R_\perp(\xi') + \frac{R_1(\xi) v^2}{1 - v^2} -  \sigma \frac{R_1(\xi') v'^2}{1 - v'^2}  + \frac{R_2(\xi)}{1 - v^2 \cos^2 \alpha} - \sigma \frac{R_2(\xi')}{1 - v'^2 \cos^2 \alpha} \Bigg] \cos \alpha & = 0,
\label{xi_eq_3} 
\end{align}
where, in turn,
\begin{align*}
R_1(\xi) = (R_\parallel(\xi) + 3 R_\perp(\xi)), \ \ \ R_2(\xi) = (R_\parallel(\xi) - R_\perp(\xi)).
\end{align*}

It can be seen from the equations (\ref{xi_eq_1} - \ref{xi_eq_3}) that at the boundary values $\alpha = 0, \ \pi/2$ only two equations remain. However, for any different non-boundary values of $\alpha$ we should consider parts in square brackets in the equations (\ref{xi_eq_2} - \ref{xi_eq_3}). To maintain the continuity of solutions, we must omit $\sin \alpha$ in (\ref{xi_eq_2}) and $\cos \alpha$ in (\ref{xi_eq_3}) for $\alpha = 0, \ \pi/2$.

\section{Critical point}

Considering the boundary cases for polar angle $\alpha$, one finds that the transverse case $(\alpha = \pi/2)$ gives
\begin{equation}
R_\parallel (\xi') - R_\perp(\xi') = \frac{R_\parallel (\xi) - R_\perp(\xi)}{\sigma}.
\label{xi2_eq}
\end{equation}
Solving the \eqref{xi2_eq} equation for $\xi'$ will always give two roots $\xi'_1 < \xi$ and $\xi'_2 > \xi$, except for the case when for $\sigma = 1$ we have $\xi' = \xi = \xi_{\textrm{crit}} (\alpha = \pi/2)$, where $\xi_{\textrm{crit}}( \pi/2) \simeq 2.62143...$ is a solution to the equation
\begin{equation}
\frac{\partial R_\parallel (\xi)}{\partial \xi} = \frac{\partial R_\perp (\xi)}{\partial \xi}.
\end{equation}

It is obvious to expect that at $\sigma = 1$ the shock wave should not exist, which corresponds to solution $v' = v, \ \xi' = \xi$. Choosing one of the two roots of the equation (\ref{xi2_eq}) for $\sigma > 1$, we want the solution $\xi' \rightarrow \xi$ when $\sigma \rightarrow 1$. Thus, the point $\xi_{\textrm{crit}}$ separates two solution spaces. If $\xi < \xi_{\textrm{crit}}$, then for the continuous limit $\sigma \rightarrow 1$ we must choose the left solution $\xi' < \xi_{\textrm{crit}}$, since only in this case the condition $\xi' \rightarrow \xi$ for $\sigma \rightarrow 1$ is satisfied. Conversely, if $\xi > \xi_{\textrm{crit}}$, then we must choose the right solution $\xi' > \xi_{\textrm{crit}}$ to ensure the same condition. If $\xi = \xi_{\textrm{crit}}$ then both solutions are possible. That is, for $\sigma > 1$ we lose the continuity of the solution $\xi'$, which shows that there is no adequate description of shock waves for arbitrary anisotropy parameter $\xi$ for the case of a fixed polar angle $\alpha' = \alpha$.

Therefore, for the limit $\xi \rightarrow 0$ one finds that left solution $\xi' \rightarrow 0$. However, for the case $\xi > \xi_{\textrm{crit}}, \ \sigma > 1$ we can not move to the isotropic limit $\xi \rightarrow 0$ since we work in another solution space. Moreover, in the case of $\xi > \xi_{\textrm{crit}}$ the solutions for the velocities provides rarefaction shock wave behavior, i.e. $v' > v$. This behavior and absence of isotropic limit clearly contradicts an adequate description of an anisotropic system. As a result, one can assume that the critical point $\xi_{\textrm{crit}}$ is an upper bound for the anisotropy parameter for the considered shock wave solutions.

Solving the system of equations (\ref{xi_eq_1} - \ref{xi_eq_3}) for polar angle $\alpha = 0$, one can find the existence of the analogously critical point $\xi_{\textrm{crit}} (\alpha = 0) \simeq 5.47941$. 

\begin{figure}[H]
\center{\includegraphics[width=0.6\linewidth]{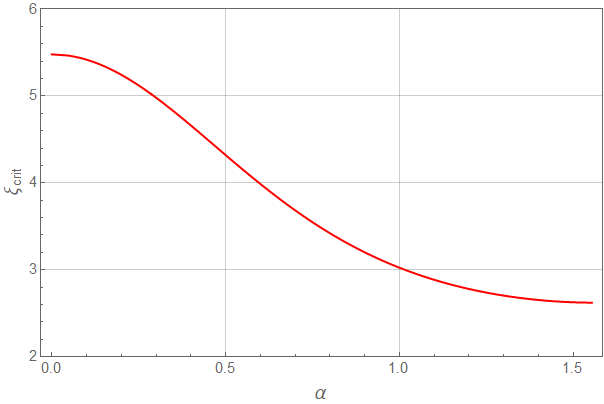}}
\caption{\small Plot of $\xi_{\textrm{crit}}$ as a function of polar angle $\alpha$.}
\label{xi_crit}
\end{figure}

The Fig. \ref{xi_crit} shows the values $\xi_{\textrm{crit}}(\alpha)$ for an arbitrary polar angle $\alpha$. The function $\xi_{\textrm{crit}}(\alpha)$ is a monotonically decreasing function that goes from $\xi_{\textrm{crit}} (\alpha = 0)$ to $\xi_{\textrm{crit}} (\alpha = \pi/2)$. For the generation of such shock waves we have a constraint condition on the anisotropy parameter $\xi$ for any polar angle $\alpha$.

The evolution of the anisotropy parameter can be obtained by solving the equations of motion both in the boost-invariant case \cite{MartStr, Martinez2012} and in the non-boost-invariant one \cite{Martinez2011}. It follows from these solutions that there may be regions where $\xi > \xi_{\textrm{crit}}$. Moreover, with the initial condition $\xi_0 = 0$, the anisotropy parameter reaches its maximum $\xi_{\textrm{max}}$ at a certain value of the proper time $\tau = \tau^*$. This maximum depends on the shear viscosity to entropy density ratio $\eta/S$. The low values of $\eta/S$ corresponds to the low values of $\xi_{\textrm{max}}$. The formation of shock waves is possible throughout the evolution of matter, if $\xi_{\textrm{max}} \leqslant \xi_{\textrm{crit}}$. For instance, in the case of purely longitudinal expansion (with initial condition $\xi_0 = 0$), the maximum value of the anisotropy parameter $\xi_{\textrm{max}} = \xi_{\textrm{crit}}(\pi/2) \simeq 2.62143$ corresponds to $\eta/S \simeq 0.17179$ and $\xi_{\textrm{max}} = \xi_{\textrm{crit}}(0) \simeq 5.47941$ corresponds to $\eta/S \simeq 0.35619$.

\begin{figure}[H]
\center{\includegraphics[width=1\linewidth]{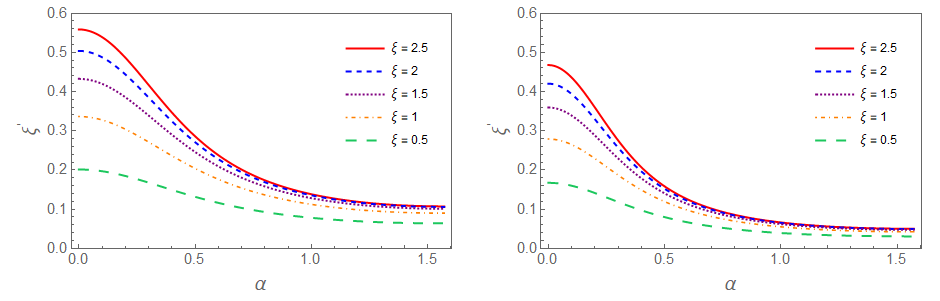}}
\caption{\small Plots of $\xi'(\alpha | \sigma)$ as a function of polar angle $\alpha$ for $\xi=2.5$ (solid), $\xi=2$ (dashed), $\xi=1.5$ (dotted), $\xi=1$ (dash-dotted) and $\xi=0.5$ (long dash) for  \(\sigma = 5\) (left) and \(\sigma = 10\) (right).}
\label{xi2_alpha}
\end{figure}

Consider the angular dependences of the anisotropy parameter $\xi'(\alpha,\sigma)$. The Fig. \ref{xi2_alpha} shows the polar angle dependence of $\xi'(\alpha | \sigma)$, where we consider the domain of $\xi \in [0, \xi_{\textrm{crit}}(\pi/2)]$. It can be seen that the system can be isotropized through the generation of shock waves. For transverse case ($\alpha = \pi/2$) we have higher anisotropy drop than for longitudinal case.

\section{Conclusion}

Solutions of shock waves were obtained in the absence of refraction of the passing flow (for polar angles we have $\alpha = \alpha'$). With a continuous change in the anisotropy parameter $\xi$ and $\sigma$ the solution for the $\xi'$ remain continuous only if we choose $\xi \leqslant \xi_{\textrm{crit}}$ or $\xi \geqslant \xi_{\textrm{crit}}$. Thus, this critical point $\xi_{\textrm{crit}}$ separates two solution spaces for shock waves. Choosing a solution at $\xi \geqslant \xi_{\textrm{crit}}$ leads to the absence of an isotropic limit $\xi \rightarrow 0$ and the behavior of shock wave solutions provide rarefaction shock wave pattern. Hence it follows that the consideration of a shock wave solutions is only possible for $\xi \leqslant \xi_{\textrm{crit}}$. Thus, the existence of shock waves is possible for small anisotropies. The connection between the evolution of the anisotropy parameter $\xi$ and the shear viscosity to entropy density ratio $\eta/S$ can also leads to the estimation of the values of $\eta/S$ at which the formation of shock waves is possible. The found values of the critical point $\xi_{\textrm{crit}}$ show that low values of $\eta/S$ are necessary for the existence of shock waves during the entire evolution of the system at zero initial anisotropy ($\xi_0 = 0$).

It was also shown that the generation of shock waves can lead to isotropization of the system since $\xi' \leqslant \xi$. The mechanism of isotropization in the case of transverse shock waves is stronger than for longitudinal shock waves.

\printbibliography

@article{Mubarak2017,
  title = {Anisotropic hydrodynamic modeling of 2.76 TeV Pb-Pb collisions},
  author = {Alqahtani, Mubarak and Nopoush, Mohammad and Ryblewski, Radoslaw and Strickland, Michael},
  journal = {Phys. Rev. C},
  volume = {96},
  issue = {4},
  pages = {044910},
  numpages = {14},
  year = {2017},
  month = {Oct},
  publisher = {American Physical Society},
  doi = {10.1103/PhysRevC.96.044910},
}

@article{Alqahtani2018,
    author = "Alqahtani, Mubarak and Almaalol, Dekrayat and Nopoush, Mohammad and Ryblewski, Radoslaw and Strickland, Michael",
    title = "{Anisotropic hydrodynamic modeling of heavy-ion collisions at LHC and RHIC}",
    eprint = "1807.05508",
    archivePrefix = "arXiv",
    primaryClass = "hep-ph",
    doi = "10.1016/j.nuclphysa.2018.10.066",
    journal = "Nucl. Phys. A",
    volume = "982",
    pages = "423--426",
    year = "2019"
}

@article{Strickland,
	author =       "Radoslaw Ryblewski, Wojciech Florkowski.",
	title =        "Anisotropic Hydrodynamics: Three Lectures",
	journal =      "Acta Physica Polonica Series",
	volume =       "B",
	number =       "45(12)",
	pages =        "2355-2394",
	year =         "2011",
	DOI =          "10.5506/APhysPolB.45.2355"
}

@article{Florkowski2013,
author = {Florkowski, Wojciech and Ryblewski, Radoslaw and Strickland, Michael},
year = {2013},
pages = {},
title = {Anisotropic Hydrodynamics for Rapidly Expanding Systems},
volume = {916},
journal = {Nuclear Physics A},
doi = {10.1016/j.nuclphysa.2013.08.004},
}

@article{Martinez2009,
  title = {Constraining relativistic viscous hydrodynamical evolution},
  author = {Martinez, Mauricio and Strickland, Michael},
  journal = {Phys. Rev. C},
  volume = {79},
  issue = {4},
  pages = {044903},
  numpages = {12},
  year = {2009},
  month = {04},
  publisher = {American Physical Society},
  doi = {10.1103/PhysRevC.79.044903},
  language = {english}
}

@article{Rischke1995,
    author = "Rischke, Dirk H. and Bernard, Stefan and Maruhn, Joachim A.",
    title = "{Relativistic hydrodynamics for heavy ion collisions. 1. General aspects and expansion into vacuum}",
    eprint = "nucl-th/9504018",
    archivePrefix = "arXiv",
    reportNumber = "CU-TP-692",
    doi = "10.1016/0375-9474(95)00355-1",
    journal = "Nucl. Phys. A",
    volume = "595",
    pages = "346--382",
    year = "1995",
  language = {english}
}

@article{Dumitru1999,
  title = {Collective dynamics in highly relativistic heavy-ion collisions},
  author = {Dumitru, A. and Rischke, D. H.},
  journal = {Phys. Rev. C},
  volume = {59},
  issue = {1},
  pages = {354--363},
  numpages = {0},
  year = {1999},
  month = {01},
  publisher = {American Physical Society},
  doi = {10.1103/PhysRevC.59.354},
  language = {english}
}

@article{Satarov2005,
author = {L.M. Satarov and H. Stocker and I.N. Mishustin},
title = {Mach shocks induced by partonic jets in expanding quark–gluon plasma},
journal = {Physics Letters B},
volume = {627},
number = {1},
pages = {64-70},
year = {2005},
issn = {0370-2693},
doi = {https://doi.org/10.1016/j.physletb.2005.08.102},
}

@article{Gyulassy1996,
    author = "Gyulassy, Miklos and Rischke, Dirk H. and Zhang, Bin",
    title = "{Hot spots and turbulent initial conditions of quark - gluon plasmas in nuclear collisions}",
    eprint = "nucl-th/9609030",
    archivePrefix = "arXiv",
    doi = "10.1016/S0375-9474(96)00416-2",
    journal = "Nucl. Phys. A",
    volume = "613",
    pages = "397--434",
    year = "1997"
}

@article{Olson1990,
title = {Plane steady shock waves in Isreal-Stewart fluids},
journal = {Annals of Physics},
volume = {204},
number = {2},
pages = {331-350},
year = {1990},
issn = {0003-4916},
doi = {https://doi.org/10.1016/0003-4916(90)90393-3},
author = {Timothy S Olson and William A Hiscock},
}

@article{MajoranaMotta,
url = {https://doi.org/10.1515/jnet.1985.10.1.29},
title = {Shock Structure in Relativistic Fluid-Dynamics},
author = {A. Majorana and S. Motta},
pages = {29--36},
volume = {10},
number = {1},
journal = {},
doi = {doi:10.1515/jnet.1985.10.1.29},
year = {1985},
}

@book{Landau,
  title={"Course of theoretical physics. Hydrodynamics"},
  author={Landau, Lev Davidovich and Lifshitz, Evgenii Mikhailovich},
  year={2013},
  publisher={Elsevier},
}

@article{Kovalenko2022,
    author = "Kovalenko A., Leonidov A.",
    title = "{Shock waves in relativistic anisotropic hydrodynamics}",
    eprint = "2103.06745",
    archivePrefix = "arXiv",
    primaryClass = "hep-ph",
    doi = "10.1140/epjc/s10052-022-10337-6",
    journal = "Eur. Phys. J. C",
    volume = "82",
    pages = "378",
    year = "2022"
}

@article{MartStr,
    author =       "Martinez, M and Strickland, M.",
    title =        "Dissipative dynamics of highly anisotropic systems",
    journal =      "Nucl.Phys.",
    volume =       "A",
    number =       "848",
    pages =        "183-197",
    year =         "2010",
    DOI =          "https://doi.org/10.1016/j.nuclphysa.2010.08.011"
}

@article{Israel,
	title = "Relativistic Theory of Shock Waves",
	author = "W. Israel",
	journal = "Proceedings of the Royal Society of London. Series A, Mathematical and Physical Sciences",
	volume = "259",
	number = "1296",
	pages = "129-143",
	year = "1960"
}

@article{Mitchell,
 author = {T. P. Mitchell and D. L. Pope},
 journal = {Proceedings of the Royal Society of London. Series A, Mathematical and Physical Sciences},
 number = {1368},
 pages = {24--31},
 publisher = {The Royal Society},
 title = {Shock Waves in an Ultra-Relativistic Fluid},
 volume = {277},
 year = {1964}
}

@article{StrRom1,
	title = "Collective Modes of an Anisotropic Quark-Gluon Plasma",
	author = "Romatschke, P. and Strickland, M.",
	journal = "Phys.Rev.",
	volume = "D",
	number = "68",
	pages = "036004",
	year = "2003",
	doi = "10.1103/PhysRevD.68.036004",
}

@article{StrRom2,
	title = "Collective modes of an Anisotropic Quark-Gluon Plasma II",
	author = "Romatschke, P. and Strickland, M.",
	journal = "Phys.Rev.",
	volume = "D",
	number = "70",
	pages = "116006",
	year = "2004",
	doi = "10.1103/PhysRevD.70.116006",
}

@article{CasalderreySolana2007,
doi = {10.1088/0954-3899/34/8/S21},
url = {https://dx.doi.org/10.1088/0954-3899/34/8/S21},
year = {2007},
month = {jul},
publisher = {},
volume = {34},
number = {8},
pages = {S345},
author = {J Casalderrey-Solana},
title = {Mach cones in quark gluon plasma},
journal = {Journal of Physics G: Nuclear and Particle Physics},
}

@article{Martinez2012,
author = {Martinez, Mauricio and Ryblewski, Radoslaw and Strickland, Michael},
year = {2012},
month = {04},
pages = {},
title = {Boost-Invariant (2+1)-dimensional Anisotropic Hydrodynamics},
volume = {85},
journal = {Physical Review C},
doi = {10.1103/PhysRevC.85.064913}
}

@article{Martinez2011,
title = {Non-boost-invariant anisotropic dynamics},
journal = {Nuclear Physics A},
volume = {856},
number = {1},
pages = {68-87},
year = {2011},
issn = {0375-9474},
doi = {https://doi.org/10.1016/j.nuclphysa.2011.02.003},
author = {Mauricio Martinez and Michael Strickland}
}

\end{document}